\documentclass[aps,prl,superscriptaddress,reprint,floatfix]{revtex4-1}
\usepackage[utf8]{inputenc}
\usepackage{amsmath}
\usepackage{amssymb}
\usepackage{graphicx}
\usepackage[colorlinks,urlcolor=blue]{hyperref}
\usepackage{url}
\usepackage{color,xcolor}
\usepackage{ulem}
\usepackage{float}
\usepackage{epstopdf}
\usepackage[none]{hyphenat}
\begin{document}

\title{Tunable Superconductivity in 1313-La$_3$Ni$_2$O$_7$: Suppressed under Compression and Possible $s^{\pm}$ Pairing under Tension}
\author{Yang Zhang}
\email{zhangy10@ornl.gov}
\affiliation{Materials Science and Technology Division, Oak Ridge National Laboratory, Oak Ridge, Tennessee 37831, USA}
\author{Ling-Fang Lin}
\email{lflin@utk.edu}
\affiliation{Department of Physics and Astronomy, University of Tennessee, Knoxville, Tennessee 37996, USA}
\author{Adriana Moreo}
\affiliation{Department of Physics and Astronomy, University of Tennessee, Knoxville, Tennessee 37996, USA}
\affiliation{Materials Science and Technology Division, Oak Ridge National Laboratory, Oak Ridge, Tennessee 37831, USA}
\author{Thomas A. Maier}
\email{maierta@ornl.gov}
\affiliation{Computational Sciences and Engineering Division, Oak Ridge National Laboratory, Oak Ridge, Tennessee 37831, USA}
\author{Elbio Dagotto}
\email{edagotto@utk.edu}
\affiliation{Department of Physics and Astronomy, University of Tennessee, Knoxville, Tennessee 37996, USA}
\affiliation{Materials Science and Technology Division, Oak Ridge National Laboratory, Oak Ridge, Tennessee 37831, USA}

\date{\today}

\begin{abstract}
Motivated by recent progress in the 1313-La$_3$Ni$_2$O$_7$ nickelate thin films (Nie et al., Nature {\bf 652}, 628 (2026)), we systematically investigate the effects of both compressive and tensile strain in this system. Self-doping effects between the single-layer (SL) and trilayer (TL) blocks are observed in our studies in both cases, but are most pronounced under tensile strain. We find that superconductivity is unlikely to emerge in the 1313 film on the LSAO substrate even with further hole doping. Remarkably, within the random-phase approximation, under {\it tensile} strain, a robust $s^{\pm}$-wave pairing state emerges in the TL subsystem with sign changes between the small electron-like $\sigma$ pocket at the $\Gamma$ point and the small hole-like $\gamma$ pocket at the M point. These pockets are connected by a wave vector close to $(\pi,\pi)$. Our calculations also suggest that superconductivity in 1313-LNO requires an optimally sized $\gamma$ pocket, because an oversized $\gamma$ pocket suppresses pairing formation. Overall, our results predict strain-dependent electronic reconstruction in 1313-La$_3$Ni$_2$O$_7$ and provide guiding principles for engineering superconductivity under ambient pressure conditions.
\end{abstract}

\maketitle

{\it Introduction--}
Since superconductivity was reported for temperatures above liquid-nitrogen in pressurized bulk bilayer (BL) La$_3$Ni$_2$O$_7$ (LNO)~\cite{Sun:arxiv}, nickelates have emerged as a central topic in the field of condensed matter physics and material science.~\cite{Zhang:arxiv-exp,Zhang:prb23,Wang:arxiv9,Dong:arxiv12,Wang:nature,Yang:arxiv09,Wang:jacs,Zhu:arxiv11,Li:cpl,Sakakibara:arxiv09,LiuZhe:arxiv,Xie:SB,Chen:arxiv2024,Zhang:prb23-2,Sakakibara:prl24,Oh:prb24,Ryee:prl,Zhang:prl1111,Braz:arxiv25,Qu:prl,Lu:prl,Liu:arxiv2023,Zhang:2323,Shi:arxiv25,Liu:arxiv,Yang:arxiv1,Liao:arxiv,Huang:arxiv,Qin:arxiv,Lin:magnetism,Maier:arxiv25,Li:nature2026,Dong:arxiv25,Zhang:2025hp,Wang:cplreview,Wang:nsr2025,Puphal:nrp2025}. The appearance of superconductivity is accompanied by a first-order structural transition from the ambient-pressure $Amam$ phase~\cite{Zhang:arxiv-exp}, containing distorted NiO$_6$ octahedra, to the high-pressure $Fmmm$ or $I4/mmm$ phase, where the octahedral tilting is suppressed~\cite{Sun:arxiv,Wang:jacs}. Remarkably, ambient-pressure superconductivity has also been reported in BL LNO thin films~\cite{Ko:nature,Zhou:nature}, which also exhibit an untilted structure~\cite{Bhatt:arxiv2025}. This discovery represents a landmark advance in nickelate superconductivity and has stimulated broad interest across the field with a focus on thin films~\cite{Wang:arxiv2502,Liu:nm25,Hao:arxiv25,Ji:arxiv25,Osada:cp25,Tarn:arxiv25,Shen:arxiv25,Li:nsr25,Wang:arxiv25,Zhang:prb26,Fan:STM,Geisler:arxiv25-1,Huang:arxiv25,Qiu:arxiv25,Cao:arxiv25,Bleys:arxiv25,Zhou:arxiv2025-60K,Liu:arxiv2026-dome,Sun:arxiv2025,Han:arxiv2026-film,Li:arxiv2025-film,Zhao:prb25,Yi:PRB2025,Geisler:arxiv25,Hua:arxiv2026,Yao:extra,Zhang:film-review,Zhang:exp-film-review}.

Interestingly, several groups simultaneously reported the synthesis of a nickelate phase with the same overall stoichiometry as BL La$_3$Ni$_2$O$_7$ (2222-LNO), namely the 1313-LNO system~\cite{Chen:jacs1313,Puphal:arxiv12,Wang:ic,Abadi:arxiv24}, featuring an alternating single-layer/trilayer (SL/TL) stacking in bulk form. More recently, bulk 1313-LNO was also reported to exhibit superconductivity above 19 GPa, albeit with a substantially lower transition temperature of 3.6 K~\cite{Huang:arxiv25-1313}, in sharp contrast to the much higher $T_c \sim 80$ K observed in high-pressure 2222-LNO bulk samples. However, another high-pressure electrical transport experiment does not reveal any discernible signatures of superconductivity up to 65 GPa; instead, they indicate pressure-induced metallization~\cite{Zhangm:1313}.

Compared with the orthorhombic $Amam$ phase of 2222-La$_3$Ni$_2$O$_7$ bulk, the ambient-pressure orthorhombic $Cmcm$ structure of 1313-LNO exhibits only weak lattice distortions. This relatively tiny structural distortion makes the 1313-La$_3$Ni$_2$O$_7$ system an ideal platform for elucidating the interplay between crystal structure and superconductivity, particularly with regards to the effects of substrate-induced strain in film format. However, in contrast to 2222-LNO films grown on LaSrAlO$_4$ (LSAO) substrates, 1313-LNO films prepared on the same substrate do not exhibit superconductivity under ambient conditions~\cite{Nie:arxiv2025-apres}. Thus, several intriguing questions naturally arise: why is ambient-pressure superconductivity absent in 1313-LNO films grown on LSAO substrates? Can 1313-LNO become superconducting in the absence of external pressure by using other substrates? If so, what is the nature of the pairing channel, and how is superconductivity linked to electronic structures?

In this work, we systematically investigate 1313-LNO films grown on substrates to address the questions raised above, by combining density functional theory (DFT)~\cite{Kresse:Prb,Kresse:Prb96,Blochl:Prb,Perdew:Prl,Mostofi:cpc} with the random-phase approximation (RPA) approaches~\cite{Kubo2007,Graser2009,Romer2020,Altmeyer2016}. We find that superconductivity is unlikely in 1313-LNO films under compressive strain on the LSAO substrate, even if Sr migration from the substrate is taken into account, consistent with recent experimental observations. However, our primary finding is that {\it superconductivity does emerge in the tensile-strain KTaO$_3$ (KTO) substrate}, with a leading $s^{\pm}$-wave pairing state from the TL subsystem. Furthermore, the outer layers in the TL blocks are antiferromagnetically coupled while the inner layer has a vanishing moment, consistent with the high-pressure bulk behavior of TL La$_4$Ni$_3$O$_{10}$.

\begin{figure}
\centering
\includegraphics[width=0.44\textwidth]{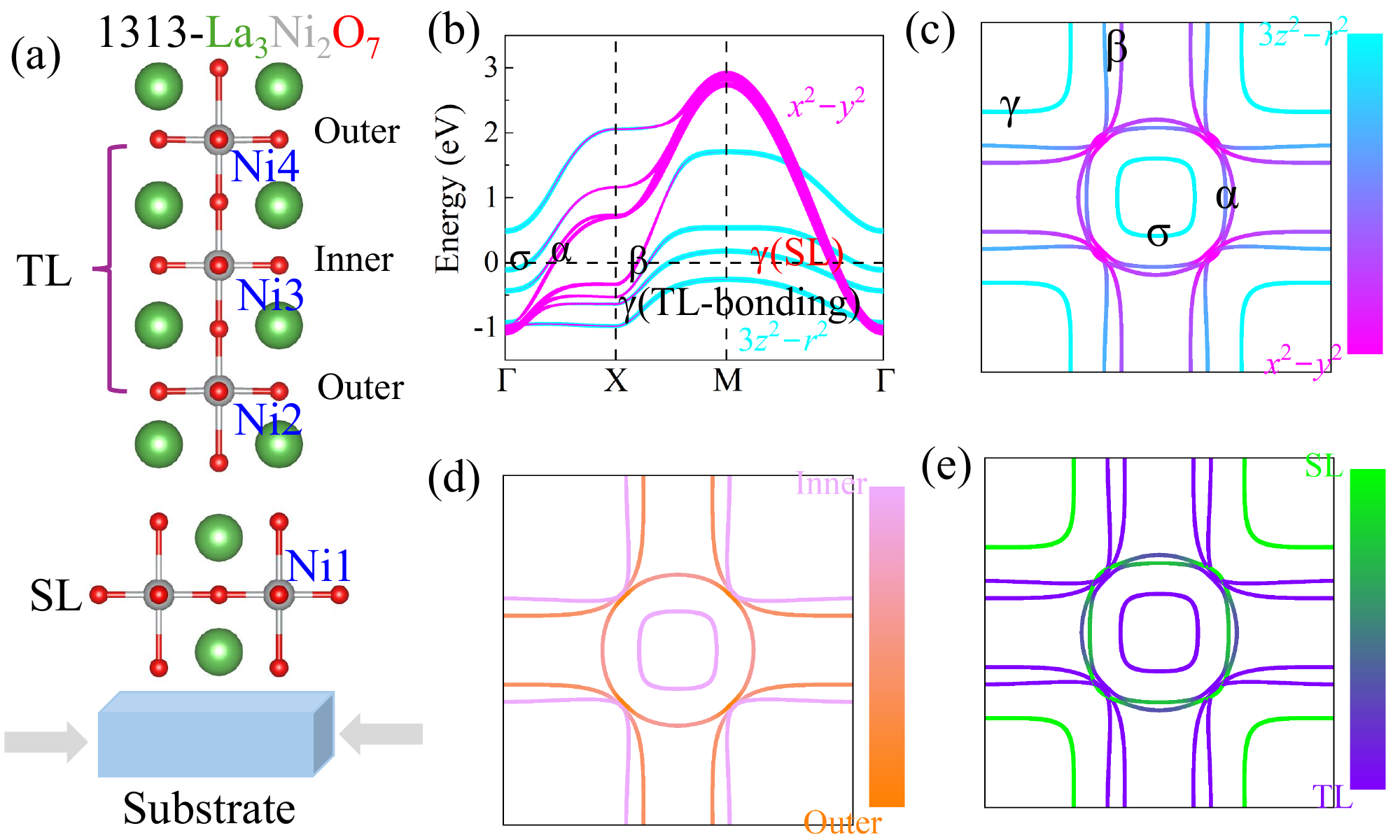}
\caption{(a) Schematic crystal structure of the 1313-LNO film on the LSAO substrate (green = La, gray = Ni, red = O), consisting of alternating SL (one Ni layer) and TL (three Ni layers) blocks, visualized using the VESTA code~\cite{Momma:vesta}. (b,c) Tight-binding band structure and Fermi surface of the 1313-LNO film on the LSAO substrate, both with total filling $n = 6$. The coordinates of the high-symmetry points of the Brillouin zone are $\Gamma$ = (0, 0, 0), X = (0, 0.5, 0), and M = (0.5, 0.5, 0). (d) Contributions of the inner and outer Ni layers to the Fermi surface of the TL subsystem, at TL filling $n = 4.19$. (e) Layer-resolved contributions of the SL and TL subsystems to the Fermi surface states at $n = 6$.}
\label{Fig1}
\end{figure}

{\it Results and Discussion--}
The 1313-LNO crystal structure features an alternating hybrid stacking of SL and TL blocks, as illustrated in Fig.~\ref{Fig1}(a), and can thus be regarded as a (La$_2$NiO$_4$)/(La$_4$Ni$_3$O$_{10}$) heterostructure along the $c$ axis. Similar to the 2222-LNO film~\cite{Ko:nature,Zhou:nature}, the LSAO substrate imposes a compressive in-plane strain ($\sim -2.6 \%$) on the 1313-LNO film, compared to the bulk at ambient conditions~\cite{Chen:jacs1313}, which consequently expands the lattice parameter along the $c$ axis~\cite{Nie:arxiv2025-apres}.

By extracting the crystal-field splittings and hopping integrals from maximally localized Wannier functions (MLWFs), obtained through mapping the DFT band structure from the MLWFs bands via the WANNIER90 package~\cite{Mostofi:cpc}, we constructed an eight-band $e_g$-orbital tight-binding model for the 1313-LNO film system on LSAO (8 bands because the unit cell contains 4 Ni's and each Ni contributes 2 orbitals). The lattice parameters were fixed at $a = b = 3.7557$~\AA\ and $c = 20.57$~\AA, as obtained in the experimental measurements of 1313-LNO films grown on the LSAO substrate~\cite{Nie:arxiv2025-apres}. The entire hopping file and computational details are provided in Supplemental Materials (SM)~\cite{Supplemental}.

As shown in Fig.~\ref{Fig1}(b), the $d_{3z^2-r^2}$ orbitals in the TL sublattice split into bonding, nonbonding, and antibonding bands, whereas no analogous splitting is observed for the SL $d_{3z^2-r^2}$ and $d_{x^2-y^2}$ orbitals. Compared to the nominal electron population of the
$e_g$ orbitals ($n = 2$ for SL La$_2$NiO$_4$ and $n = 4$ for TL La$_4$Ni$_3$O$_{10}$), the calculated electronic densities are approximately 1.81 and 4.19 for the SL and TL subsystems, respectively. These results indicate the presence of a ``self-doping'' effect, characterized by charge transfer from the SL to the TL sublattice, thereby {\it electron-doping the TL subsystem}.

In addition, the $\gamma$-band associated with the TL $d_{3z^2-r^2}$ bonding state does {\it not} cross the Fermi level, leading to the absence of the $\gamma$ pocket shown in Fig.~\ref{Fig1}(c). This conclusion is in agreement with recent angle-resolved photoemission (ARPES) experiments for 1313-La$_3$Ni$_2$O$_7$ thin films grown on the LSAO substrate~\cite{Nie:arxiv2025-apres}. Moreover, the splitting of the $\beta$ sheets originates from multilayer splitting between the inner and outer layers of the TL sublattice, arising from slight differences in the on-site energies, as shown in Fig.~\ref{Fig1}(d).

To elucidate the absence of superconductivity observed experimentally~\cite{Nie:arxiv2025-apres}, we next investigate the superconducting properties of the 1313-LNO film on the LSAO substrate by performing multi-orbital RPA calculations for the coupled SL-TL tight-binding model, based on a perturbative weak-coupling expansion in these Coulomb interaction terms~\cite{Kubo2007,Graser2009,Altmeyer2016,Romer2020}. The pairing strength $\lambda_\alpha$ and the gap structure $g_\alpha({\bf k})$ for channel $\alpha$ are obtained by solving an eigenvalue problem of the form
\begin{eqnarray}\label{eq:pp}
	\int_{FS} d{\bf k'} \, \Gamma({\bf k -  k'}) g_\alpha({\bf k'}) = \lambda_\alpha g_\alpha({\bf k})\,.
\end{eqnarray}
The momenta ${\bf k}$ and ${\bf k'}$ are restricted to the Fermi surface, and the singlet pairing interaction $\Gamma({\bf k- k'})$ is given by the irreducible particle-particle vertex. Within RPA, the dominant term entering $\Gamma({\bf k-k'})$ is the RPA spin susceptibility $\chi({\bf k-k'})$. More computational details of RPA are provided in SM~\cite{Supplemental}

The RPA gives a spin-density-wave (SDW) state for the SL subsystem already for small $U$, driven by the near-perfect nesting of the $\gamma$-sheet Fermi surface states [see green regions in Fig.~\ref{Fig1}(e)], similar to previous studies on the 1313-La$_3$Ni$_2$O$_7$ high-pressure bulk systems~\cite{Zhang:1313,LaBollita:1313,Lechermann:PRM2024,Chen:arxiv26}. This is also supported by recent ARPES experiments for 1313-LNO single crystals, where SL is gapped with an SDW state~\cite{Au-Yeung:NP}. Other previous studies also suggested that the observed superconductivity in 1212-La$_5$Ni$_3$O$_{11}$ originates primarily from the BL subsystem, while the SL block mainly serves as a bridge for interlayer Josephson coupling~\cite{Zhang:arXiv2506-1212,Shao:nc26}. We therefore use a restricted TL-only model, in which the Coulomb matrix elements on the SL subsystem are set to zero, similar to the restricted BL-only model we used in our previous study of 1212-La$_5$Ni$_3$O$_{11}$~\cite{Zhang:1212}, to prevent the SL subsystem from directly participating in the pairing.

\begin{figure}
\centering
\includegraphics[width=0.44\textwidth]{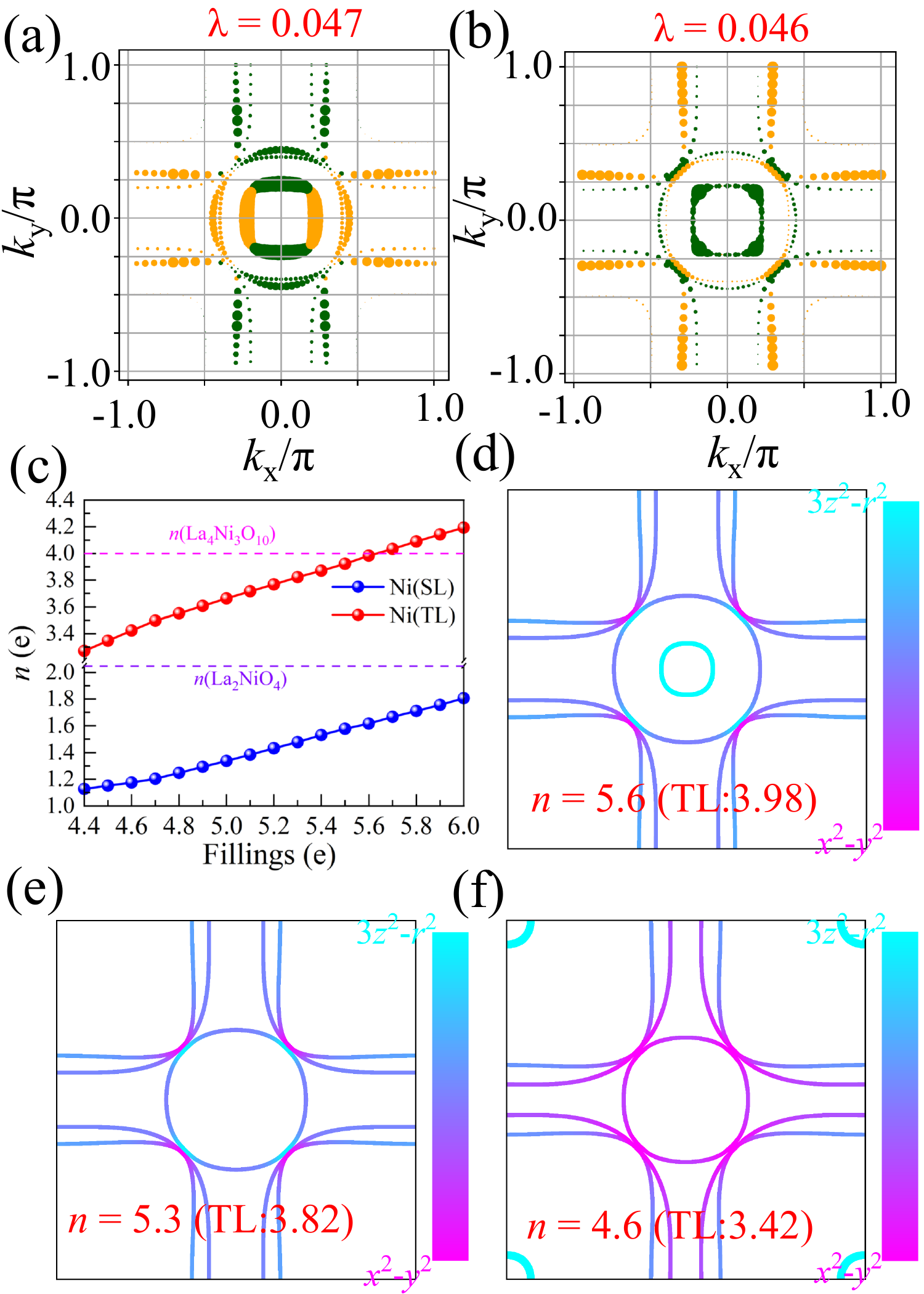}
\caption{(a-b) The RPA-calculated leading superconducting singlet gap structures $g_\alpha(\mathbf{k})$ on the Fermi surfaces are shown for (a) the $d_{x^2-y^2}$-wave channel  with eigenvalue $\lambda = 0.047$ and for the (b)$s^{\pm}$-wave channel with eigenvalue $\lambda = 0.046$, for 1313-La$_3$Ni$_2$O$_7$ films grown on an LSAO substrate for $U = 1.2$, $U'=U/2$, $J=J' = U/4$, calculated for temperature $T=0.01$~eV considering the TL-restricted model. Here, the SL states are included in the Hamiltonian, but the interactions $U$, $U'$, $J$, $J'$ are set to zero on the SL orbitals. Thus, the SL subsystem will not directly participate in the pairing since the spin-fluctuation interaction for the SL orbitals will be zero. The sign of $g_\alpha({\bf k})$ is indicated by the color (orange = positive, blue = negative), and its amplitude by the point size. (c) Total electron occupations of the SL and TL sublattices as a function of the overall filling $n$ in the tight-binding model. (d,e,f) Orbital contributions at the Fermi surface from the trilayer subsystem for different overall fillings $n$.}
\label{Fig2}
\end{figure}

By solving the eigenvalue problem of Eq.~(\ref{eq:pp}) within the RPA for this TL-restricted model, we find that the leading pairing instability occurs in the $d_{x^2-y^2}$-wave channel, with a competing $s^{\pm}$-wave state with smaller eigenvalue $\lambda$. However, increasing $U$ further, the value of the coupling $\lambda$ remains relatively small, even when close to the critical $U_c$ [see results for $U = 1.2$ eV as an example, in Figs.~\ref{Fig2}(a) and (b)]. This behavior is similar to the evolution of the RPA eigenvalue $\lambda$ with the effective interaction $U$ in the BL La$_3$Ni$_2$O$_6$ system~\cite{Zhang:prb24}. It indicates that pairing correlations are weak in this system and that a possible low-temperature superconducting instability is likely preempted by a magnetic instability in the stoichiometric 1313-LNO films grown on an LSAO substrate.

Considering the buffer-layer structure observed in ARPES results~\cite{Nie:arxiv2025-apres}, as well as previous studies suggesting possible Sr migration from the substrate into the nickelate layer in bilayer thin films~\cite{Li:nsr25}, we also consider the possibility of hole doping induced by Sr migration in this 1313-LNO film on the LSAO substrate. As shown in Fig.~\ref{Fig2}(c), the SL Ni1 site loses more electrons per Ni than other sites in the TL sublattice upon hole doping. In addition, with increasing hole doping, the inner-layer $\sigma$ pocket originating from the $d_{3z^2-r^2}$ orbital of the TL sublattice is gradually suppressed [see Fig.~\ref{Fig2}(d)] and disappears at a total filling of approximately $n = 5.3$ (1.325 electrons per Ni) shown in Fig.~\ref{Fig2}(e), corresponding to $17.5\%$ hole doping on the Ni site ($11.67\%$ Sr$^{2+}$ substitution on the La$^{3+}$ sites).

Notably, this small $\sigma$ pocket around the $\Gamma$ point was not observed in ARPES measurements of the 1313-LNO film grown on the LSAO substrate~\cite{Nie:arxiv2025-apres}, whereas it was detected in another ARPES study of the bulk compound under ambient pressure~\cite{Abadi:arxiv24}. This observation suggests that the 1313-LNO film may be hole-doped, which may be caused by the Sr migration from the LSAO substrate. As hole doping increases further, the $\gamma$ pocket from the lowest TL bonding state is induced around $n = 4.6$ ($23.33\%$ Sr$^{2+}$ doping), as displayed in Fig.~\ref{Fig2}(f). Similar to our previous study of the high-pressure TL model~\cite{Zhang:arxiv24}, the pairing strength is strongly suppressed once either the $\sigma$ or $\gamma$ pocket disappears. Consequently, superconductivity is unlikely to emerge in the 1313-LNO film on the LSAO substrate, even in the presence of Sr migration.

\begin{figure}
\centering
\includegraphics[width=0.44\textwidth]{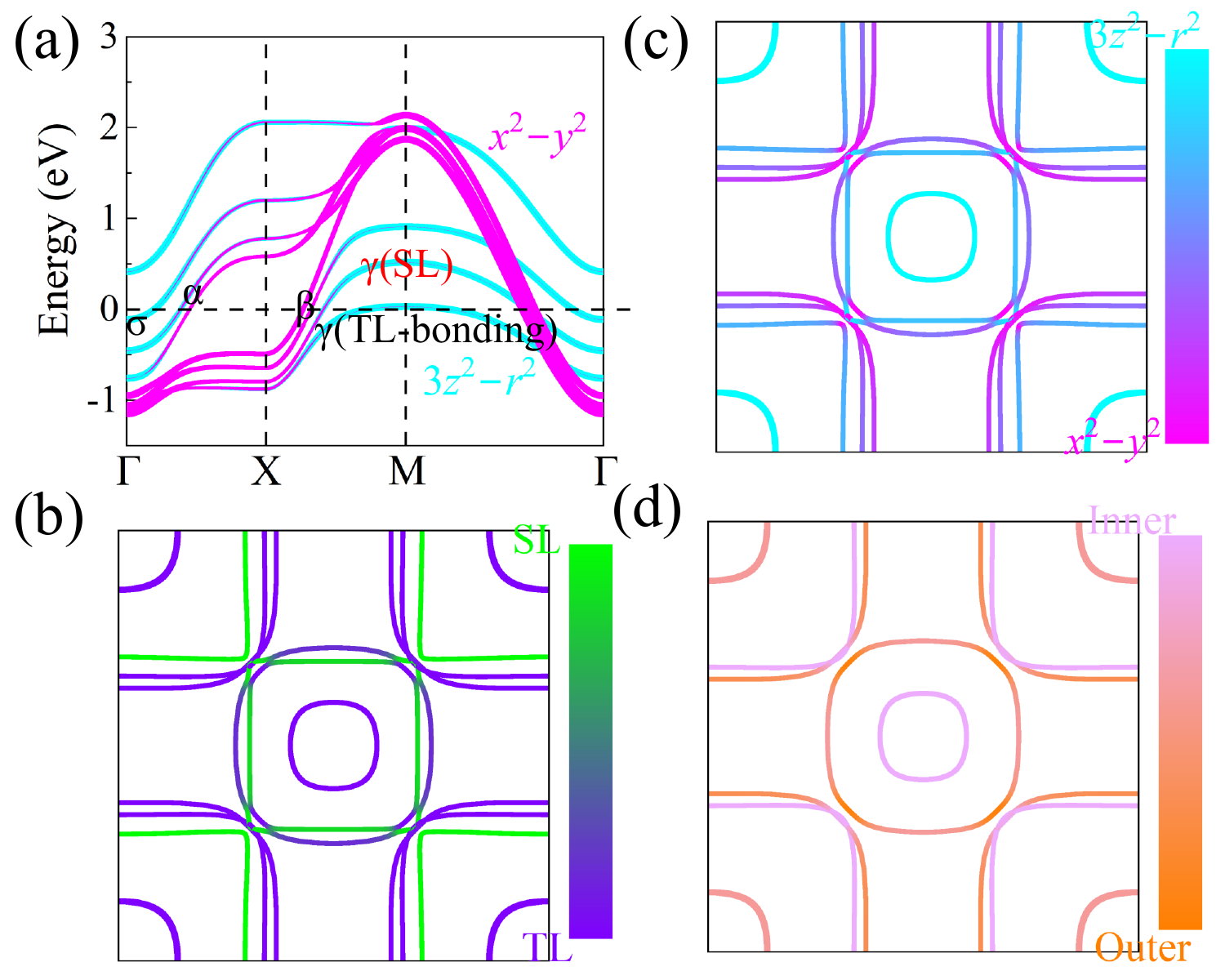}
\caption{(a-b) Tight-binding band structure and Fermi surface for the 1313-La$_3$Ni$_2$O$_7$ film on the KTO substrate, with total filling $n = 6$, respectively. (c) Layered contributions of the SL and TL subsystems to the Fermi-surface states for a total filling $n = 6$. (d) The contributions of inner and outer layers to the Fermi surface of the trilayer sublattices, at TL filling $n = 4.38$. The hopping parameters are provided in SM~\cite{Supplemental}.}
\label{Fig3}
\end{figure}

We noticed that by introducing a tensile-strain substrate, the out-of-plane lattice spacing is reduced, and this may stabilize the key $\gamma$ pocket, which could in turn favor superconductivity. Motivated by this possibility, we next investigate 1313-LNO grown on tensile-strain KTO $(\sim 3.4 \%)$ to explore potential superconducting behavior. Remarkably, the TL $\gamma$ bands of the $d_{3z^2-r^2}$ bonding states now cross the Fermi level, leading to the appearance of the $\gamma$ pocket around the M point, as shown in Figs.~\ref{Fig3}(a) and (b). Moreover, the bandwidth of 1313-LNO on the KTO substrate is reduced by approximately $24\%$ compared to the LSAO case, as the tensile strain suppresses the in-plane hopping amplitudes of the $d_{x^2-y^2}$ orbitals. Accordingly, the ratio of the interlayer $d_{3z^2-r^2}$ hopping to the intralayer $d_{x^2-y^2}$ hopping increases to approximately 1.5, compared with $\sim 1.3$ in the LSAO case. Moreover, the in-plane hybridization between the $d_{x^2-y^2}$ and $d_{3z^2-r^2}$ orbitals is also enhanced in the KTO case compared with the 1313-LNO film grown on the LSAO substrate.

In addition, the self-doping effect is further enhanced in the KTO case, with an increased electron transfer from the SL to the TL blocks, where the corresponding calculated electronic occupancies are approximately 1.62 and 4.38 for the SL and TL subsystems, respectively. Furthermore, Fig.~\ref{Fig3}(c) shows the layer-resolved Fermi-surface contributions from the SL and TL parts, which are qualitatively similar to those obtained for the LSAO substrate.

The Fermi surface topology of the TL part closely resembles that of high-pressure TL bulk La$_4$Ni$_3$O$_{10}$~\cite{Zhang:arxiv24}. Analysis of the layer-resolved contributions to the TL Fermi surface in Fig.~\ref{Fig3}(d) reveals that the $\sigma$ pocket around the $\Gamma$ point, predominantly derived from the inner layers, exhibits favorable nesting with the emergent $\gamma$ pocket around the M point, which mainly originates from the outer layers. In this scenario, superconductivity with a sign-changing order parameter between these two pockets is expected. Motivated by this consideration, we performed an RPA pairing analysis of a restricted TL-only model for the 1313-La$_3$Ni$_2$O$_7$ system grown on the KTO substrate, once again assuming that the SL subsystem does not directly participate in the pairing.

\begin{figure}
\centering
\includegraphics[width=0.44\textwidth]{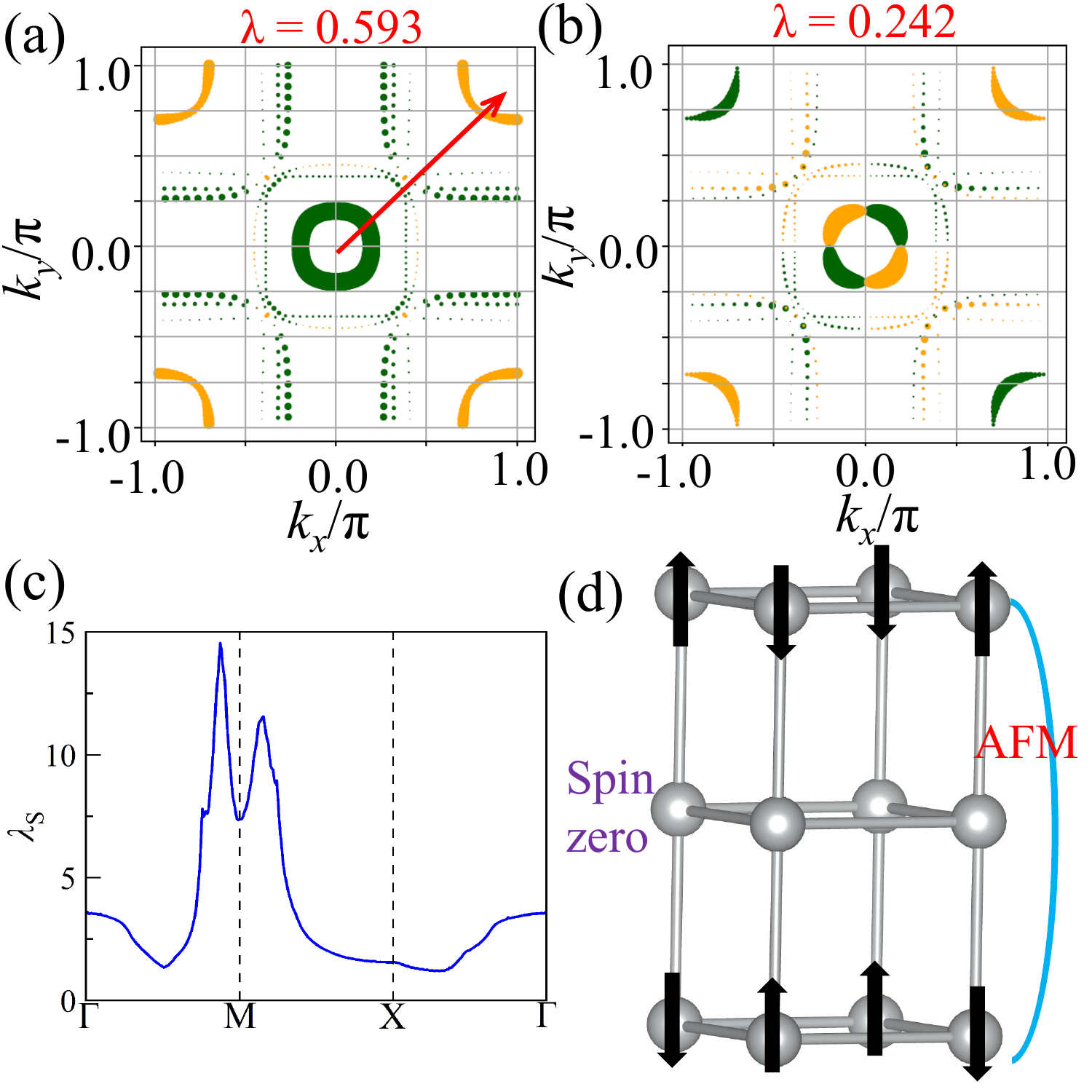}
\caption{(a-b) The RPA-calculated superconducting singlet gap structures $g_\alpha(\mathbf{k})$ on the Fermi surfaces for (a) the leading solution $s^{\pm}$-wave state with eigenvalue $\lambda = 0.593$ and (b) subleading $d_{xy}$-wave channel with eigenvalue $\lambda = 0.242$, for 1313-La$_3$Ni$_2$O$_7$ films grown on the KTO substrate. Here we considered the TL-restricted model (as used in Figs.~\ref{Fig2}(a-b)) with $U=0.9$~eV, $U'=U/2$, $J=J'=U/4$, and $T=0.01$~eV in the TL subsystem and $U=U'=J=J'=0$ in the SL block. The sign of $g_\alpha({\bf k})$ is indicated by the color (orange = positive, green = negative), and its amplitude by the point sizes. (c) Leading eigenvalue $\lambda^s_0({\bf q})$ of the spin susceptibility tensor $\chi^s_{\ell_1\ell_1\ell_2\ell_2}({\bf q})$, along a high-symmetry path in the Brillouin zone.
This calculation is performed in a mixed representation with in-plane ($k_x$, $k_y$) dependence and in the real-space domain for the out-of-plane dependence, where the real-space eigenvector at $q_{\rm peak}$ reveals the out-of-plane SDW structure. (d) Schematic of the magnetic spin density structure of the TL sublattice obtained from RPA, with spin-up and spin-down marked by black arrows. The outer layers are coupled antiferromagnetically, while the middle layer has vanishing spin density, similar to high-pressure TL La$_4$Ni$_3$O$_{10}$ bulk~\cite{Zhang:arxiv24}. For simplicity,
we show ${\bf q} = (\pi, \pi)$ for in-plane magnetic correlation.}
\label{Fig4}
\end{figure}

Figures~\ref{Fig4}(a) and (b) show the gap structures of the leading solution $s^{\pm}$-wave state and subleading $d_{xy}$-wave channel that are obtained from solving the eigenvalue problem in Eq.~(\ref{eq:pp}). The leading $s^{\pm}$-wave instability with robust eigenvalue $\lambda = 0.593$ is well separated from the subleading $d_{xy}$-wave solution with eigenvalue $\lambda = 0.242$. As shown in Fig.~\ref{Fig4}(a), the superconducting order parameter of the leading $s^{\pm}$-wave state changes sign between the small electron-like $\sigma$ pocket at the $\Gamma$ point and the small hole-like $\gamma$ pocket at the M point, which are connected by a wave vector close to $(\pi,\pi)$.  We note that the pairing strength of the leading $s^{\pm}$-wave channel, $\lambda = 0.593$, is significantly larger than that obtained for 2222-LNO at 25 GPa ($\lambda = 0.290$ with the same interaction strength $U = 0.9$~eV but calculated at a much lower temperature $T=0.001$~eV~\cite{Zhang:nc24} vs. $T=0.01$~eV used here). However, this does not necessarily translate to a higher $T_c$. Since the pairing found here is confined to the TL blocks, which are separated by the SL blocks, a three-dimensional superconducting state requires Josephson tunneling of the pairs in the TL blocks across the SL blocks, analogous to the mechanism proposed for the 1212 nickelates~\cite{Zhang:arXiv2506-1212,Shao:nc26}. Nevertheless, our calculations suggest that superconductivity with a potentially robust $T_c$ could be achievable in 1313-LNO films grown on the KTO substrate.

To understand the origin of the leading $s^{\pm}$-wave pairing instability, we also examine the structure of the RPA spin susceptibility tensor $\chi({\bf q})$ for 1313-LNO films grown on the KTO substrate, which is obtained from the Lindhard function $\chi_0({\bf q})$ as $\chi({\bf q}) = \chi_0({\bf q})[1-{\cal U}\chi_0({\bf q})]^{-1}$. Here, all the quantities are rank-four tensors in the orbital indices $\ell_1, \ell_2, \ell_3, \ell_4$ and ${\cal U}$ is a tensor involving the interaction parameters~\cite{Graser2009}.
The physical spin susceptibility is obtained by summing the pairwise diagonal parts of the tensor, $\chi_{\ell_1\ell_1\ell_2\ell_2}({\bf q})$, over $\ell_1$, $\ell_2$.

Figure~\ref{Fig4}(c) shows the leading eigenvalue $\lambda^s_0({\bf q})$ of $\chi_{\ell_1\ell_1\ell_2\ell_2}({\bf q})$ along the high-symmetry directions in the Brillouin zone. A pronounced peak ${\bf q}$ emerges near the M point, indicating strong in-plane antiferromagnetic (AFM) incommensurate correlations that are close to $(\pi,\pi)$. This spin state can be understood in terms of the competition between intraorbital and interorbital hopping mechanisms, as discussed in previous studies~\cite{Lin:prl21,Lin:cp,Lin:prb22}. The corresponding eigenvector for ${\bf q}$ at the peak corresponds to a spin density wave state with antiparallel moments on the outer layers of the TL part, while the middle layer exhibits zero spin density, as schematically illustrated in Fig.~\ref{Fig4} (d). In addition, the eigenvector also shows that the AFM correlation between the outer layers of the TL blocks mainly arises from the $d_{3z^2-r^2}$ orbitals. This magnetic configuration closely resembles that of high-pressure TL-La$_4$Ni$_3$O$_{10}$ in the bulk~\cite{Zhang:arxiv24}.

\begin{figure}
\centering
\includegraphics[width=0.44\textwidth]{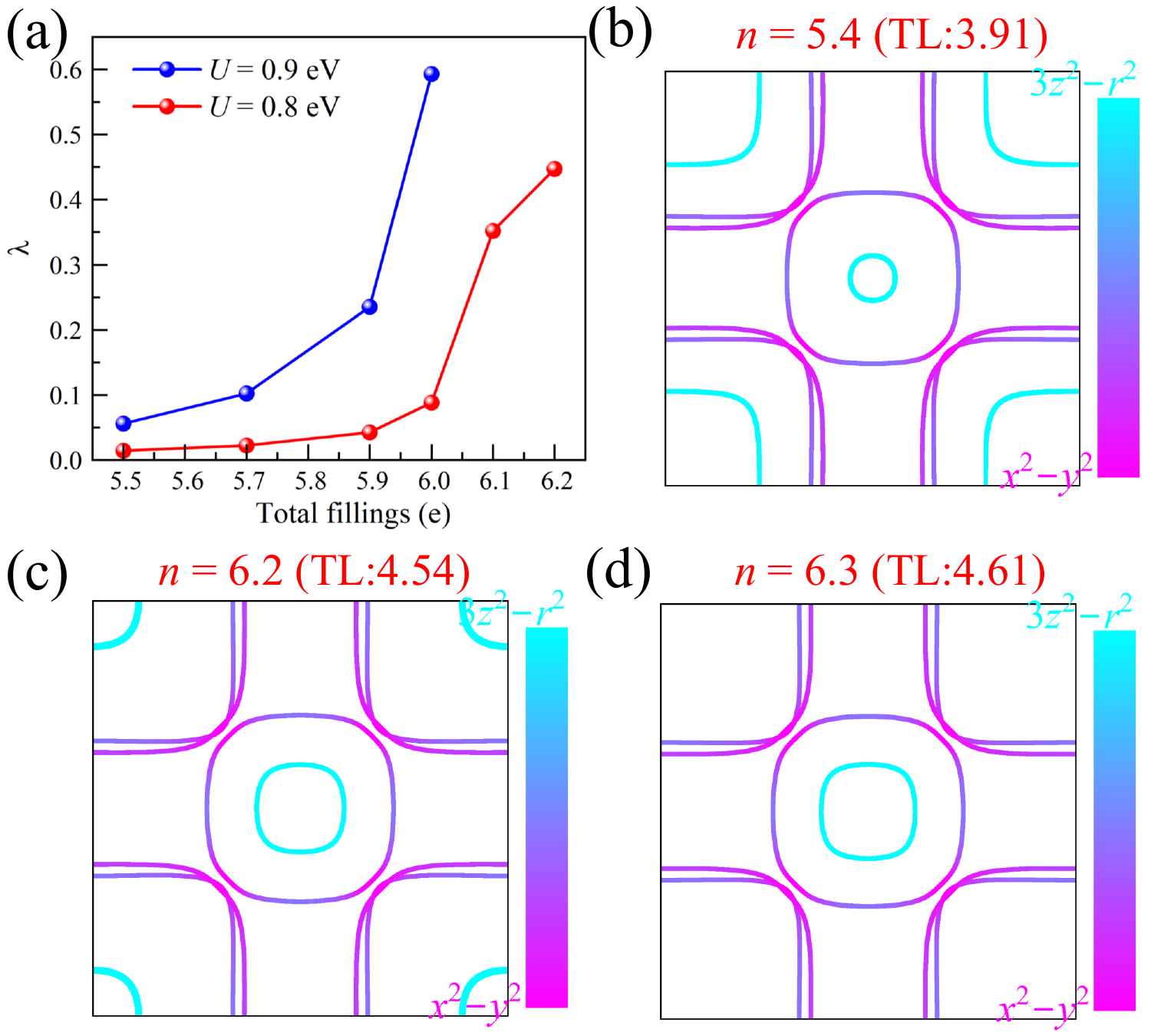}
\caption{(a) The RPA calculated pairing strength $\lambda$ for the $s^{\pm}$ instability for various total electronic densities for 1313-LNO on KTO. Here we used again the TL-restricted model with $U=0.9$~eV, $U'=U/2$, $J=J'=U/4$, and $T=0.01$~eV in the TL subsystem and $U=U'=J=J'=0$ in the SL block. (b-d) Tight-binding Fermi surface for the TL part of 1313-LNO on the KTO substrate: (b) $n = 5.4$; (c) $n = 6.2$; (d) $n = 6.3$.}
\label{Fig5}
\end{figure}

To understand the evolution of the pairing instability with doping, we further investigated the doping dependence of the RPA pairing strength $\lambda$. We find that the $s^{\pm}$ pairing channel remains the leading instability throughout the entire doping range we studied. As shown in Fig.~\ref{Fig5}(a), hole doping continuously suppresses $\lambda$. With increasing hole concentration, the size of the central $\sigma$ pocket decreases, whereas that of the $\gamma$ pocket increases. Near $n = 5.4$ (corresponding to approximately $10 \%$ 2+ substitution of La$^{3+}$), the pairing strength is strongly suppressed, even though the TL block still possesses both $\gamma$ and $\sigma$ pockets, as illustrated in Fig.~\ref{Fig5}(b).

Under electron doping, the calculated pairing strength $\lambda$ increases, indicating an enhancement of the superconducting pairing instability. Here, to further clarify the trend shown in Fig.~\ref{Fig5}(a), we also present results obtained with a lower interaction strength, $U = 0.8$ eV, since $U = 0.9$ eV already exceeds the critical value $U_c$ for spin density wave formation at $n = 6.2$. Note that electron doping at the La$^{3+}$ site is challenging for nickelates because stable 4+ substitutional ions are relatively rare. For increasing the electron doping, the size of the central $\sigma$ pocket increases, but that of the $\gamma$ pocket decreases (see Fig.~\ref{Fig5}(c) as an example). Since the discussed $s^{\pm}$-wave pairing originates from the nesting between the central $\sigma$ pocket and the $\gamma$ pocket around the M point, the superconducting pairing is suppressed above $n = 6.3$, where the $\gamma$ pocket (Fig.~\ref{Fig5}(d)) is absent.

Thus, we find that hole doping suppresses pairing, while electron doping provides a strong boost to pairing. Our calculations also suggest that the $\gamma$ pocket is essential for superconductivity in 1313-LNO (or related TL systems), but superconductivity is suppressed again if this pocket becomes too large.

{\it Conclusion--}
In summary, we have systematically investigated the effects of compressive and tensile strain on 1313-LNO films. The SL subsystem exhibits an instability driven by near-perfect nesting between the $\alpha$ and $\gamma$ Fermi-surface sheets. In addition, we find strain-tunable superconductivity in the 1313-LNO system: the compressive-strain LSAO substrate suppresses superconductivity, in agreement with recent experiments, even when Sr migration is considered. In contrast, a leading $s^{\pm}$-wave pairing state emerges from the TL subsystem in the tensile-strain KTO film, consistent with superconductivity in high-pressure bulk TL La$_4$Ni$_3$O$_{10}$. Moreover, while additional hole doping weakens the pairing instability, electron doping is found to boost pairing in 1313-LNO on KTO. Our calculations also suggest that superconductivity in 1313-LNO requires an optimally sized $\gamma$ pocket, because an oversized pocket suppresses pairing. Overall, these results provide clear design guidelines for 1313-LNO systems and suggest a viable route toward realizing superconductivity under ambient pressure conditions.

{\it Acknowledgment--}
This work was supported by the U.S. Department of Energy, Office of Science, Basic Energy Sciences, Materials Sciences and Engineering Division.  This manuscript has been authored by UT-Battelle, LLC, under contract DE-AC05-00OR22725 with the US Department of Energy (DOE). The US government retains, and the publisher, by accepting the article for publication, acknowledges that the US government retains a nonexclusive, paid-up, irrevocable, worldwide license to publish or reproduce the published form of this manuscript, or allow others to do so, for US government purposes. DOE will provide public access to these results of federally sponsored research in accordance with the DOE Public Access Plan (https://www.energy.gov/DOE Public Access Plan).

{\it Data availability--}
The dataset of the main findings of this study is openly available in the Zenodo Repository. In addition, the hopping and crystal-field parameters for our tight-binding and RPA calculations are available in a separate file of the Supplementary Material and Zenodo Repository for reproducing our results. Simulation RPA codes are available at https://github.com/maierta/MRPAPP.

\end{document}